\documentclass[12pt]{article}
\usepackage{latexsym}
\usepackage{amssymb}
\usepackage{amsfonts}

\usepackage{color,graphicx} 
\definecolor{brownn}{cmyk}{0,1,1,0.5}
\definecolor{bluen}{rgb}{.1 ,0, .8}

\usepackage{ifpdf}
\ifpdf
\RequirePackage[T1]{fontenc}
\usepackage[hypertexnames=false,
colorlinks,linkcolor=bluen,citecolor=brownn,urlcolor=brownn]{hyperref}
\usepackage{epstopdf} 
\else
\usepackage[hypertex,hypertexnames=false,backref,
colorlinks,linkcolor=bluen,citecolor=brownn,urlcolor=brownn]{hyperref}
\fi
\graphicspath{{images/}}

\def\nc#1{\newcommand{#1}}
\def\rnc#1{\renewcommand{#1}}

\def\a{\alpha}
\def\b{\beta}
\nc{\g}{\gamma}
\def\d{\delta}
\nc{\D}{\Delta} 
\nc{\e}{\eta}
\nc{\ep}{\epsilon}

\nc{\ve}{\varepsilon}
\nc{\G}{\Gamma}

\nc{\la}{\lambda}
\nc{\La}{\Lambda}
\nc{\om}{\omega}
\nc{\Om}{\Omega}
\nc{\vphi}{\varphi}
\nc{\si}{\sigma}
\nc{\Si}{\Sigma}
\rnc\th{\theta}
\nc\Th{\Theta}
\nc{\z}{\zeta}

\nc{\got}[1]{\mathfrak{#1}} 

\nc\im{{\rm Im}\, }
\nc\re{{\rm Re}\, }

\def\Tr{{\rm Tr}}

\nc{\Rt}{{\tilde R}}
\nc{\CC}{{\mathbb C}}
\nc\II{{\mathbb I}} 
\nc{\RR}{{\mathbb R}}
\nc{\HH}{{\mathbb H}}
\nc{\NN}{{\mathbb N}}
\nc{\ZZ}{{\mathbb Z}}
\nc{\MM}{{\mathbb M}}

\nc{\ov}[1]{\overline{#1}}

\nc{\non}{\nonumber\\}
\def\nn{\nonumber}
\nc{\noi}{\noindent}

\nc{\p}{\partial}
\nc{\na}{\nabla}

\nc\vev[1]{\ensuremath{\lan #1\ran} {}}
\nc\refeq[1]{(\ref{#1})}
\nc{\eqref}[1]{(\ref{#1})}
\rnc\to[1][]{\ensuremath{\stackrel{#1}{\rightarrow\;}}}

\nc{\twovec}[2]{\left( \!\!
\begin{array}{c} #1\\  #2 \end{array}\!\!\right)}
\nc{\twomat}[4]{\left(\!\! \begin{array}{cc} #1&#2\\ 
#3&#4\end{array}\!\! \right)}

\nc{\ds}{\displaystyle}

\nc{\lan}{\langle}
\nc{\ran}{\rangle}

\nc{\beq}{\begin{equation}}
\nc{\eeq}{\end{equation}}
\nc{\beqa}{\begin{eqnarray}}
\nc{\eeqa}{\end{eqnarray}}
\nc{\beqas}{\begin{eqnarray*}}
\nc{\eeqas}{\end{eqnarray*}}
\nc{\barr}{\begin{array}}
\nc{\earr}{\end{array}}
\nc{\ben}{\begin{enumerate}}
\nc{\een}{\end{enumerate}}
\nc{\bit}{\begin{itemize}}
\nc{\eit}{\end{itemize}}
\rnc\-{\item}

\nc{\cred}{\color{red}}
\nc{\cblue}{\color{blue}}

\nc\rQ[1][]{\ensuremath{{\cred\leftarrow(?)\mbox{\footnotesize #1}} } }

\nc\more{{ \cred{MORE}}}
\nc{\remark}[1]{\cblue[Rem]\footnote[*]{\color{blue}{Remark:} #1}}
\nc{\foot}[1]{{}{\cblue[{}\footnote[0]{\tt #1}]}}
\nc{\ADD}{{\cblue ADD}}

\nc\fb{\ensuremath{{\bar 5}} {}}
\nc\yfb{\ensuremath{Y_{\overline 5}} {}}
\nc\yf{\ensuremath{Y_5} {}}
\nc\yt{\ensuremath{Y_{10}} {}}
\nc\ytb{\ensuremath{Y_{\overline{10}}} {}}
\nc\y[2][]{\ensuremath{Y_{#2}^{#1}} {}}
\nc\f[2][]{\ensuremath{5_{(#2)}^{#1}} {}}
\nc\hfb[1][]{\ensuremath{{\bar 5}^{\, #1}_H} }
\nc\yu[2]{\ensuremath{y_{#2}^{(#1)}}{}}

\definecolor{lightgray}{cmyk}{0.1,0.2,0,0.1}

\nc\lgut{\La_{GUT}}
\nc\mpl{M_{Pl}}
\nc\fdx{r_{D/X}}

\normalsize

\begin{document}
\thispagestyle{empty}
\title{Masses and FCNC in Flavoured GMSB scheme}
\author{ T. Jeli\'nski$^{1}$, J. Pawe{\l}czyk$^{2}$\\
{\small {}}\\
{\small \it $^{1}$Department of Field Theory and Particle Physics, Institute of Physics, University of Silesia,}\\
{\small \it Uniwersytecka 4, 40-007 Katowice, Poland}\\
{\small {\it $^{2}$Institute of 
Theoretical Physics, Faculty of Physics, University of Warsaw,}}\\{\small {\it 
Ho\.za 69, 00-681 Warsaw, Poland}}\\
}
\date{}
\maketitle
\abstract{\normalsize

We discuss a specific model of extended GMSB type in which \fb component of one messenger vector pair   is treated as an extra flavour. FCNC effects related to the superpotential couplings between \fb messenger and $i$-th family of MSSM matter, $h_i$, are analysed. We find that the strongest limit, $h_1h_2\lesssim10^{-3}$, emerges from 
constraints on RR slepton contribution to radiative muon decay. 
The results are in agreement with simple models of hierarchical Yukawa interactions.

}

\section{Introduction}
\normalsize

LHC puts strong constraints on many BSM models. Among them MSSM and its extensions  are still very promising candidates. 
 Close to TeV scale the models contain plethora of free parameters,
 the so-called soft terms.
For any theoretical model   the latter cannot be
 arbitrary but depend on its UV completion. 
Soft terms are strongly bound by the experimental data.
LHC provided  limits on masses of SUSY particles while the lack of excesses in the FCNC processes constrains possible mixings between different flavours of sfermions. These have been discussed for many years, see e.g. \cite{Gabbiani:1996hi,Masiero:2005ua,Isidori:2013ez,Calibbi:2012yj} and references therein, and the limits are constantly improved \cite{Arana-Catania:2013nha,Arana-Catania:2014ooa}. The important data comes from low energy physics of mesons and leptons.

In this paper, we are going to discuss soft terms 
%
generated in one of so-called Flavoured GMSB models \cite{Shadmi:2011hs, Albaid:2012qk, Abdullah:2012tq, 
Calibbi:2013mka, Galon:2013jba}, in which messengers couple to all families of MSSM matter. The model under consideration 
has the only messengers in $5+\fb$ of $SU(5)_{\mathrm{GUT}}$ and \fb will be treated as an ordinary extra flavour.
 As a consequence  all the flavours and the messenger $\overline{5}$'s must  have
common Yukawa 
couplings which we expect to posses natural hierarchical structure.
This simple assumption plus some natural requirements, quite surprisingly, put strong constraints on possible extra couplings.
The very few 
new parameters of the model will be constrained analysing FCNC 
and LHC data.
In the discussion,  we shall use as often as possible analytical results which give better feeling of various derived bounds.
This will be supplemented by numerical analysis using various packages.
It is interesting to see that off-diagonal soft masses does not change significantly with RG flow.

The structure of the paper is the following.
In the next section, we shall recall main features of the model under consideration. 
The Section $3$ is devoted to analytical, leading loop analysis of the soft terms generated through the process of SUSY breaking and its mediation.
Next, in the Section $4$, we shall discuss dependence of the soft masses on the strongest 
messenger-flavour coupling $h_3$. After that, we use
FCNC processes to bound other parameters of the model. 
Final section provides discussion of the obtained results.


\section{The model of Flavoured GMSB}

Let us briefly describe the model. The motivation behind the construction is an observation that both flavour visible chiral matter and messengers  may originate from the same 
D7-brane intersection in some F-theory constructions of GUT theories (the so-called F-GUTs) \cite{HV-E8}.
It is also believed that hierarchical structure of Yukawa couplings is generated by some non-perturbative physics: either by fluxes \cite{Yukawa-HV} or non-trivial condensate on a distant brane \cite{Yukawa-np} and it has Froggatt-Nielsen type structure \cite{Froggatt-Nielsen}. 
Thus all these fields should have common Yukawa matrix. 
The model proposed here contain one extra multiplet of \fb of GUT $SU(5)$ which will have common Yukawa coupling matrix with the standard three families. Anomaly cancellation is provided by additional $5$ which will form the messenger field.
Perturbativity of couplings implies that the most obvious model of this type can contain only one vector pair of messengers and  $\tan\b$ cannot be bigger than 10 \cite{jp-yukawa}.

We begin with writing down all  relevant superpotential terms using $SU(5)_{\mathrm{GUT}}$ representations at the GUT scale:\footnote{Here we use common notation for Yukawa couplings for down-quarks and leptons.}
\beq\label{mainY}
\sum_{i,j}y^u_{ij}10_i10_j (5_H)_2+
\sum_{i,J}\hat y_{iJ}10_i\fb_J (\fb_H)_2+(\sum_J a_J \fb_J) 5 X,
\eeq
 where the subscript 2 attached to the Higgs fields recalls that we take into account only the doublet   part of  the 5 and $\overline{5}$ of Higgses.
We let the flavour indices run as follows: $i,j=1,2,3;\; J=1,...4$ and there is also an extra $5$ which will form the messenger vector pair with one of the $\overline{5}$'s. 
All the couplings in \refeq{mainY} have hierarchical structure i.e.
$y^u_{i,j}\ll y^u_{i+1,j}$ etc. We stress that also $a_J\ll a_{J+1}$.
Moreover we expect that the couplings with the top-most indices $y^u_{33},\ \hat y_{34}$ and $a_4$ to be of the order 1.
The spurion vev $\vev X$  gives mass $M_Y$ to the messengers (and  triggers SUSY breaking through its nontrivial $F$-term), one of which is a mixture of basic fields: $\yfb\sim \sum_J a_J \fb_J $ and leaves the other $\overline{5}$'s of matter massless. These are ordinary flavour fields. 
In order to decouple the messengers we need to diagonalize K\"ahler potential\footnote{This was discussed in \cite{Shih-soft-m}.} and go to the mass eigenstate basis. This results in  the following Yukawa matrix:
\beq\label{y34}
\hat y = \left(
\barr{cccc}
y_{1} & 0& 0& h_1\\
0&y_2&  0& h_2\\ 
0&0& y_{3}& h_3
\earr
   \right).
\eeq
Hierarchical structure assumed above can now be rephrased as
$y_1\ll y_2\ll y_3\ll h_3\sim 1$ and $h_1\ll h_2\ll h_3$.
Of course, the precise relations between couplings could look differently for the quarks $(\bar 3,1/3)\in\fb$ (e.g. $h_1\sim \ep^2,\ h_2\sim \ep^2,\ h_3\sim 1 $)
and leptons $(2,-1/2)\in\fb$ (e.g. $h_1\sim \ep^4,\ h_2\sim \ep^2,\ h_3\sim 1,\ \ep\sim 0.2$) \cite{Yukawa-HV}. For the  simplicity of the notation,  
 we have decided to suppress these differences here.

Couplings $h_i$'s heavily influence properties of soft terms which we shall analyse in the course of the paper. In short,  $h_1$ and $h_2$ induce FCNC processes and will be limited from above by FCNC, while $h_3$ will mainly influence mass spectra of the supersymmetric particles. 

\section{Soft terms}
Supersymmetry is broken by non trivial $F$-term, $F_X$, of a spurion superfield $X$ which vev \vev X also provides masses for messengers.
 As discussed in the previous sections, we focus on the case of $\tan\beta\lesssim10$. 

All soft masses at $M_Y=\vev X$ depend on $\xi=F_X/\vev X$ only. The scale $M_Y$ appears only through RG evolution. In the following, we shall adjust $\xi$ and $M_Y$ to respect all the known phenomenological constraints. The latter are mainly the mass of the lightest Higgs boson $h^0$, $m_{h^0}\approx126\,\mathrm{GeV}$ and the bound for the gluino mass $M_{\widetilde{g}}\gtrsim 1.7\cdot 10^3\,\mathrm{GeV}$ \cite{Aad:2014wea}.  

Below the scale $M_Y$, messengers decouple leaving three massless flavour families and bunch of soft terms of which masses will  depend on $h_i$'s and gauge coupling constants $\alpha_r$. General formulae were derived in \cite{Shih-soft-m,Chacko:2001km} and it is enough to adopt them for the case at hand.
Soft masses of sfermions at the scale $M_Y$ are:
\beqa\label{sm-u}
	\label{sm-q}
(m_{\widetilde Q})^2_{ij}&=&\frac{\xi ^2}{3840 \pi ^4}\,[\d_{ij}\ 8 \pi ^2 \left(\alpha _1^2+45 \alpha _2^2+80 \alpha_3^2\right)\non
&&\qquad\qquad+h_i^*h_j\, (105 \left| h_3\right| {}^2-4 \pi  \left(7
   \alpha _1+45 \alpha _2+80 \alpha _3\right)+\D_{ij}^Q)],\\
(m_{\widetilde U})^2_{ij}&=&\frac{\xi^2}{30 \pi^4}\, [\d_{ij}\ \pi^2 
\left(\alpha_1^2+5 \alpha_3^2\right)
-\d_{i3}\d_{\!j\,3}\, \frac{15}{64}\,|y^u_{3}|^2|h_3|^2 ],\\
	\label{sm-d}
(m_{\widetilde D})^2_{ij}&=&\frac{\xi ^2}{120 \pi ^4}\,[\d_{ij}\ \pi ^2 \left(\alpha
   _1^2+20 \alpha _3^2\right)
	-\d_{i3}\d_{j3}\,\frac{75}{16}\, |y^d_{3}|^2
   \left| h_3\right|{}^2 ],\\
	\label{sm-l}
(m_{\widetilde L})^2_{ij}&=&
\frac{3\xi ^2}{160 \pi ^4}\,[\d_{ij}\ \pi ^2 
\left(\a_1^2+5 \a_2^2\right)
	-\d_{i3}\d_{\!j\,3}\,\frac{5}{4}\, |y^e_{3}|^2  |h_3|{}^2],\\
	\label{sm-e}
	(m_{\widetilde E})^2_{ij}&=&\frac{\xi ^2}{640 \pi ^4}\,[\d_{ij}\ 48 \pi ^2 \alpha_1^2+h_i^*h_j\,(35 \left|\, h_3\right| {}^2-12 \pi  \left(3
   \alpha _1+5 \alpha _2\right)+\D_{ij}^E)].
\eeqa
The formulae (\ref{sm-q}-\ref{sm-e}) were obtained using some simplifications.
First of all, it appears that RG flow between $M_{\mathrm{GUT}}\approx 10^{16}$ GeV (including mixings between all $\fb$'s) and the messenger scale $M_Y$, which we assume to be bigger than $10^{8}$ GeV, 
has very mild influence on the structure and values of Yukawa couplings and the couplings $h_i$'s. Thus, it is save to say that (\ref{sm-q}-\ref{sm-e})
are effective terms defined at $M_Y$.
Secondly, they contain only leading terms in $h_1,\ h_2$'s and 
$y_1,\ y_2$'s with noticeable exception of left squarks and right sleptons where next-to-leading contributions ($\D$'s) might be important due to possible cancellation between $h_3$ and $\alpha_r$ contributions. This will be discussed in the following sections. The important $\D$'s for FCNC processes are:
\beqa
\D^Q_{12}&=&45|y^d_3|^2+15|y^e_3|^2+105(|h_1|^2+|h_2|^2),\\
\D^Q_{13}&=&75|y^d_3|^2+15|y^e_3|^2+105(|h_1|^2+|h_2|^2),\\
\D^E_{12}&=&15|y^d_3|^2+5|y^e_3|^2+35(|h_1|^2+|h_2|^2),\\
\D^E_{13}&=&15|y^d_3|^2+10|y^e_3|^2+35(|h_1|^2+|h_2|^2). 
\eeqa
Soft masses for Higgses are:
\beqa
m_{H_{u}}^{2}&=&\frac{3\xi^{2}}{160\pi^{4}}\left[\pi^{2}(\alpha_{1}^{2}+5\alpha_{2}^{2})-\frac{5}{8}|h_{3}|^{2}|y^{u}_{3}|^{2}\right],\\
m_{H_{d}}^{2}&=&\frac{3\xi^{2}}{160\pi^{4}}\left[\pi^{2}(\alpha_{1}^{2}+5\alpha_{2}^{2})\right.
\nonumber\\
&&\qquad\quad\left.+\frac{1}{24}|h_{3}|^{2}\left(140|h_{3}|^{2}+15|y^{u}_{3}|^{2}-16\pi(4\alpha_{1}+15\alpha_{2}+20\alpha_{3})\right)\right],\nn
\eeqa
 and the trilinear terms:\footnote{They enter the scalar potential in the following way: $V\supset-H_{u}\widetilde{Q}_{i}(T_{u})_{ij}(\widetilde{u}_{R}^{*})_{j}+H_{d}\widetilde{Q}_{i}(T_{d})_{ij}(\widetilde{d}_{R}^{*})_{j}+H_{d}\widetilde{L}_{i}(T_{e})_{ij}(\widetilde{e}_{R}^{*})_{j}$ (SLHA2 conventions \cite{Allanach:2008qq}). $A$-term $A_{t}$ is related to the $(T_{u})_{33}$: $A_{t}=(T_{u})_{33}/y^{u}_{33}$.}
\beqa\label{trilinear}
(T_{u})_{ij}&=&-\frac{\xi}{16\pi^{2}}h_{i}h_{k}^{*}y^{u}_{kj},\nn\\
(T_{d})_{ij}&=&-\frac{\xi}{16\pi^{2}}\left[4h_{k}h_{k}^{*}y^{d}_{ij}+h_{i}h_{k}^{*}y^{d}_{kj}\right],\\
(T_{e})_{ij}&=&-\frac{\xi}{16\pi^{2}}\left[4y^{e}_{ij}h_{k}h_{k}^{*}+2y^{e}_{ik}h_{k}^{*}h_{j}\right].\nn
\eeqa

At one loop gaugino masses at $M_Y$ are given solely by values of gauge coupling constant, e.g. gluino mass is $M_{\widetilde{g}}=\xi\a_{3}/4\pi$. In the following, we shall often express masses of the SUSY particles in terms of $M_{\widetilde{g}}$.

Apparently, some of the soft masses (\ref{sm-q}-\ref{sm-e}) become negative for some $h_3$ rising questions about stability of the MSSM potential. 
The naive bounds obtained with all $\a$'s equal 1/20 at $M_Y$, $y^u_3\approx0.6$ and $\tan\b<10$ are 
\beq
h_3<\frac1{12}\quad\mbox{or}\quad \frac14<h_3<1.3.
\eeq
This behaviour will be reflected in numerical analysis yielding
excluded regions in $h_3$ space.

\section{Mass spectrum}

Here we shall explore influence of messenger-flavour Yukawa couplings on mass spectrum of the supersymmetric particles.  
It is clear that the most relevant coupling is $h_3$.
We perform numerical analysis of the model using appropriately changed codes of  \texttt{SuSpect} \cite{Djouadi:2002ze} and \texttt{SPheno} \cite{Porod:2003um,Porod:2011nf}. Higgs bosons sector shall be studied with the help of 
\texttt{FeynHiggs} \cite{Heinemeyer:1998yj,Heinemeyer:1998np,Degrassi:2002fi,Frank:2006yh,Hahn:2013ria}. 

Scanning over $0\leq h_{3}\leq1.2$ reveals the following constraints. Values of messenger coupling in the range $0.25\lesssim h_3\lesssim 0.75$ or $h_3\gtrsim1.0$ give not acceptable phenomenology due to either tachyons or charge/colour breaking. This corresponds to a rough estimate of the potential stability given in the previous section. For the remaining part of the tested range of $h_3$, mass spectrum can be characterised as follows.

Mass of the lightest Higgs boson,  
 $m_{h^0}$, hardly depends on $h_3$, what is a consequence of relatively small coefficient 
in the 1-loop contribution to $A_t$-term induced by messenger coupling. Although both $A_t$ and stop masses $m_{\overline{t}_{1,2}}$ do depend on $h_{3}$ (see discussion below), the stop mixing parameter $|X_t/M_S|$ does not change much when one varies $h_3$. 
Hence, to accommodate for $m_{h^0}\sim123-127\,\textrm{GeV}$ in the model under consideration, one has to adjust $\xi$ scale at least to $5\cdot 10^5\,\textrm{GeV}$. As a consequence, the gluino mass is at least $M_{\widetilde{g}}=3.4\cdot10^3\,\textrm{GeV}$.

The remaining part of Higgs boson sector i.e. $H^0$, $A^0$ and $H^{\pm}$ are sensitive to $h_3$, but their masses are always bigger than $0.4M_{\widetilde{g}}$. Masses of lighter neutralinos $\widetilde{\chi}_{1,2}^0$ are $(0.20,0.38)M_{\widetilde{g}}$ respectively while $\widetilde{\chi}_{3,4}^0$ are heavier than $0.65M_{\widetilde{g}}$. Lighter  chargino mass is $0.38M_{\widetilde{g}}$, while mass of $\widetilde{\chi}_2^{\pm}$ is bigger than $0.65M_{\widetilde{g}}$. 

\begin{figure}[!h]
\begin{center}
\includegraphics[scale=0.9]{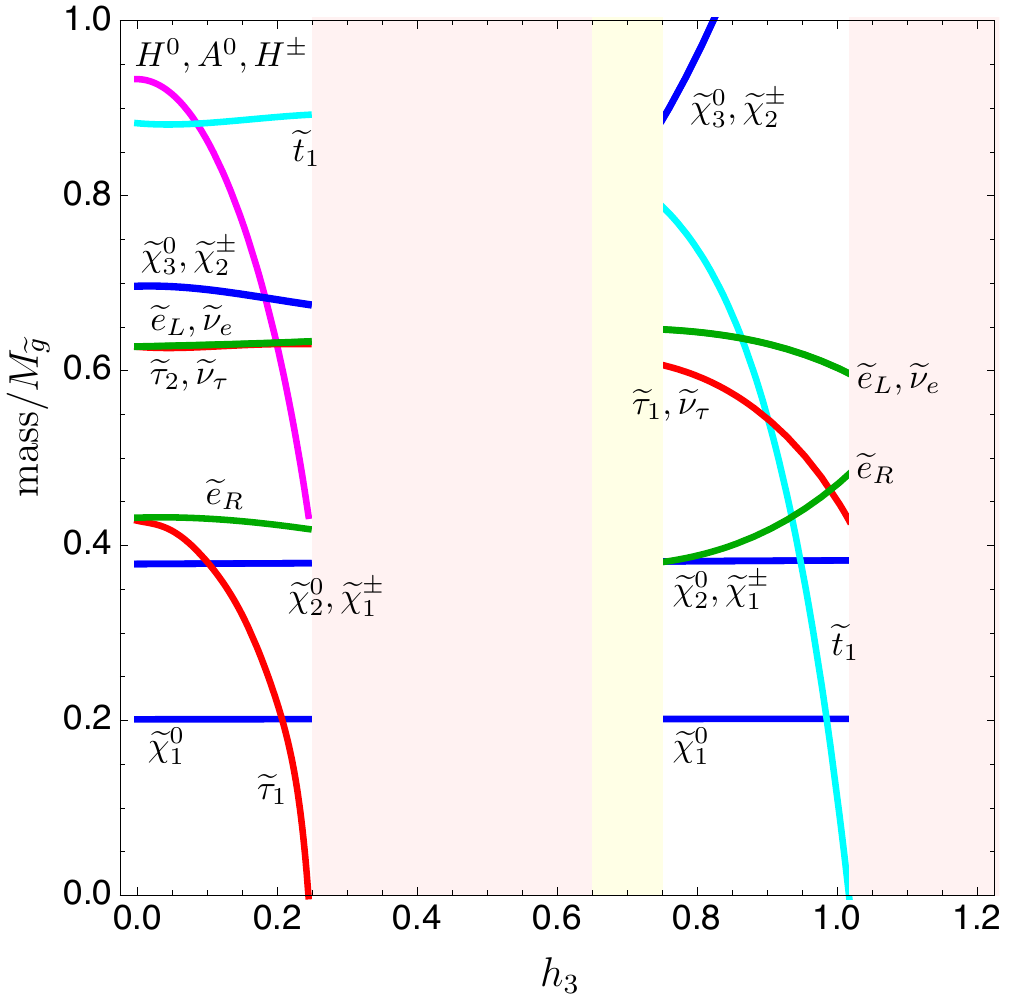}
\caption{
Masses of the sparticles vs. messenger coupling $h_3$; $M_{\widetilde{g}}=3.4\cdot10^3\,\mathrm{GeV}$, $\tan\beta=10$,  while messenger scale is set to $M_Y=10^{14}\,\mathrm{GeV}$. Colour code: $\widetilde{\tau}$ (red), $\widetilde{e}$ (green), $\widetilde{t}$ (light blue), $\widetilde{\chi}$ (dark blue), $H^0,A^0,H^{\pm}$ (magenta). The pink vertical bands  show regions of $h_3$ for which one gets tachyons,
while the yellow band represents region for which charge/colour breaking occurs.
}
\label{spectrum}
\end{center}
\end{figure}

Squarks of the 1st and 2nd generation are almost insensitive to $h_3$. For all tested values of $h_3$ their masses are bigger than $1.1M_{\widetilde{g}}$, hence they lie well above ATLAS exclusion limit \cite{Aad:2014wea}. On the other hand, messenger coupling $h_3$ is relevant for the 3rd generation of squarks. Increasing value of $h_{3}$ heavily changes values of their soft terms at $M_Y$, what eventually results in lowering mass of $\widetilde{t}_{1}$ and increasing mass of $\widetilde{t}_{2}$ (see Fig. \ref{spectrum}). For $h_{3}\gtrsim0.95$ the lighter stop mass can be as low as $0.3M_{\widetilde{g}}$. For such a case, $\widetilde{t}_{1}$ is the NNLSP and it decays mainly in the channel $\widetilde{t}_1\rightarrow t\widetilde{\chi}_1^0$. When approaching $h_3=1.0$, first, masses of $\widetilde{t}_1$ and $\widetilde{\chi}^0_1$ become degenerate and then $\widetilde{t}_1$ becomes the NLSP with mass approximately $0.2\,M_{\widetilde{g}}$ or even less.  Such light $\widetilde{t}_1$ is still not excluded by the LHC \cite{Atlas-2013-024} because $\widetilde{\chi}^0_1$ is rather heavy,  $m_{\widetilde{\chi}_1^0}=684\,\mathrm{GeV}$.   
Messenger coupling only slightly influences sbottoms masses - for all values of $h_3$ they are heavier than $M_{\widetilde{g}}$.  
%

It turns out that $h_3$ is also relevant for masses of lightest sleptons. Increasing $h_{3}$ from $0$ to $0.25$ drives lighter stau mass to values as small as $0.1M_{\widetilde{g}}$.  
 Here $\widetilde{\tau}_1$ is mostly right-handed and it can be NLSP or NNLSP depending on the precise value of $h_3$ (see Fig. \ref{spectrum}). For $h_3$ bigger than $0.25$ lighter stau becomes tachyonic. Its $(\textrm{mass})^{2}$ starts to be positive again for $h_{3}\gtrsim0.5$. Because there is an overlap of that region with region excluded due to tachyonic $H^0,A^0,H^{\pm}$ or  charge/colour breaking, one finds phenomenologically acceptable spectrum only for $h_{3}\gtrsim0.75$. In that region, the lighter stau mass is also lowered by messenger coupling, but now $\widetilde{\tau}_1$  is mostly left-handed. Finally, let us comment on the mass pattern of the remaining sleptons. They are only slightly altered by messenger coupling and for all values of $h_3$ their masses are bigger than $0.35M_{\widetilde{g}}$.   

It is worth to note that the two aforementioned allowed ranges of $h_3$ correspond to so different mass spectra. One can check that in the first `window' either the lightest neutralino $\widetilde{\chi}^0_1$, which is bino-like, or the lighter stau $\widetilde{\tau}_1$ can be the NLSP. On the other hand, in the second `window' either $\widetilde{\chi}_1^0$ or lighter stop $\widetilde{t}_1$ play the role of the NLSP.

\begin{figure}[!h]
\begin{center}
\includegraphics[scale=0.55]{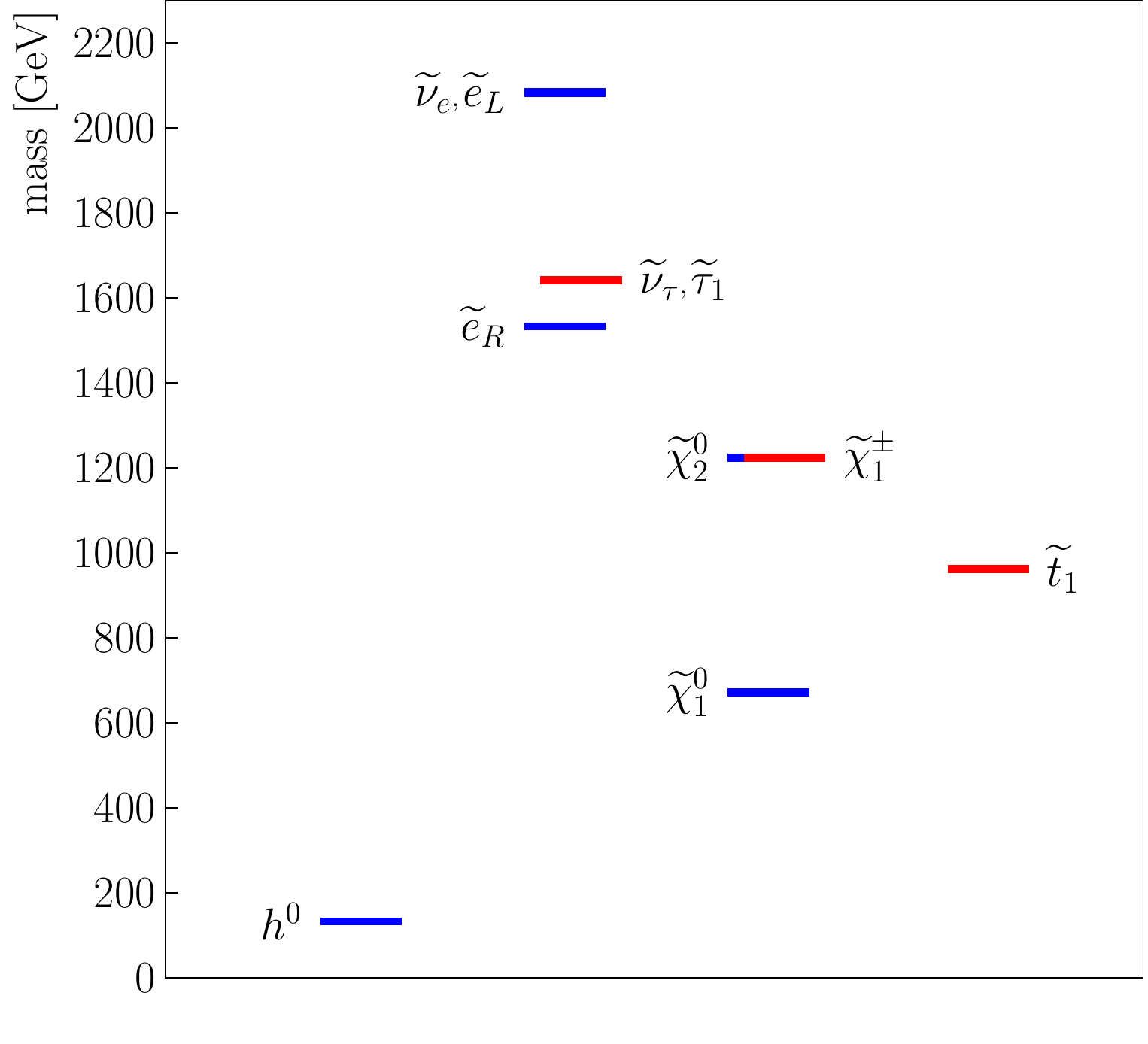}
\caption{
Spectrum of the model for $h_3=0.97$. Masses of the sparticles which are not displayed on the plot are bigger than $3.4\cdot10^3\,\mathrm{GeV}$.  
The plot has been made with the help of \texttt{PySLHA} \cite{Buckley:2013jua}.
}
\label{spectrum1}
\end{center}
\end{figure}

\section{Constraints from flavour physics}

In the models under consideration, messengers couple to all families of MSSM matter what generates non-diagonal soft terms. Hence,  squarks and sleptons $(\textrm{mass})^{2}$ matrices have non-trivial flavour structure enhancing contributions to various FCNC processes.
Below, we shall discuss bounds on $h_1$ and $h_2$ coming from this source. 

Here  we extend our discussion to cover the whole region 
$0<h_3<1.2$. The reason is that in some modifications of the discussed model there can also be present additional contributions to diagonal soft masses coming e.g. from higher dimensional operators $X^\dagger X \Phi^\dagger\Phi$, ($\Phi$ being any matter field) generated by exchange of an extra $U(1)$ heavy gauge boson. This would alleviate the problem of potential stability not changing the flavour structure. 

In order to analyse 
contributions to FCNC processes, it is convenient to follow \cite{Gabbiani:1996hi} and
parameterize squarks and sleptons $(\textrm{mass})^2$ matrices at the EWSB scale $Q=\sqrt{m_{\widetilde{t}_1}m_{\widetilde{t}_2}}$ by $(\delta^{f}_{AB})_{ij}^{(Q)}$:
\beq\label{M2}
(M^2_{f})_{ij}^{(Q)}=\overline{M}_{f}^{2(Q)}\left(\begin{array}{cc}(\delta_{LL}^{f})_{ij}^{(Q)}&(\delta_{LR}^{f})_{ij}^{(Q)}\\(\delta_{RL}^{f})_{ij}^{(Q)}&(\delta_{RR}^{f})_{ij}^{(Q)}\end{array}\right),
\eeq
where ${\overline{M}^{2(Q)}_{f}}=\Tr(M_{f}^{2(Q)})/6$ is the average mass of squarks ($f=u,d$) or sleptons ($f=e$). 
Taking into account GMSB contributions to soft terms these masses  can be approximated as:
\beq\label{m2av}
{\overline{M}^{2(M_Y)}_{u,d}}=\frac{\alpha_{3}^{2}}{6\pi^{2}}\xi^{2},\qquad{\overline{M}^{2(M_Y)}_{e}}=\frac{15\alpha_{2}^{2}}{256\pi^{2}}\xi^{2}.
\eeq
Numerical check shows that $\overline{M}^{2(Q)}_{u,d}$  only weakly depends on $h_3$ and it differs from\footnote{In \eqref{m2av} we set $\a_{2,3}=1/20$.} \eqref{m2av} by a factor $1.5$. In the slepton case, to get proper value of $\overline{M}^{2(Q)}_{e}$ one has to correct \eqref{m2av} by a factor 0.8 for $h_3\lesssim0.2$, while for $h_3\gtrsim0.75$ messenger coupling $h_3$ substantially increases average mass of sleptons such that \eqref{m2av} has to be multiplied by a factor $1.5-4.0$ (depending on the value of  $h_3$).

Moreover, off-diagonal soft terms of \eqref{sm-q}-\eqref{trilinear} also do not change the order of magnitude 
by the RG flow and the rotation to the SCKM basis (i.e. diagonalization of Yukawa matrices). Altogether,  
 $(\delta_{AB}^{f})_{ij}^{(Q)}\approx (\delta_{AB}^{f})_{ij}^{(M_Y)}$ up to factors of the order a few. 
Hence, the main features of the elements 
$(\d^{f}_{AB})_{ij}^{(Q)}$, $i\neq j$, can be estimated using their values at the messenger scale $M_Y$.
This helps to have an analytical control over FCNC corrections.
In the formulae below, these numerical factors have been taken into account. 
 
 Below, we shall discuss limits on the couplings $h_1,\ h_2$ entering  $(\delta^{f}_{AB})_{ij}^{(Q)}$, $i\neq j$.
We shall use
mass insertion approximation (MIA) \cite{Gabbiani:1996hi}
known to catch basic characteristic of contributions to FCNC processes.
In the MIA,  non-diagonal $(\delta^{f}_{AB})_{ij}^{(Q)}$ terms are treated as additional interactions.
This will be cross-checked with direct numerics. With the help of appropriately modified code of \texttt{SPheno-3.3.0}, we have validated the dependence of the FCNC observables on messenger couplings without referring to MIA.

As an experimental input we shall use the following $\Delta F=2$ processes: meson mixing $D^{0}-\overline{D}^{0}$ ($f=u$, $i=1$, $j=2$), $\overline{K}^{0}-K^{0}$ ($f=d$, $i=1$, $j=2$), $\overline{B}_{d}^{0}-B_{d}^{0}$ ($f=d$, $i=1$, $j=3$), $\overline{B}_{s}^{0}-B_{s}^{0}$ ($f=d$, $i=2$, $j=3$) and $\Delta F=1$ processes: $b$ quark decay $b\rightarrow s \gamma$ ($f=d$, $i=2$, $j=3$) and radiative lepton decays $\tau\rightarrow \mu \gamma$ ($f=e$, $i=2$, $j=3$), $\tau\rightarrow e\gamma$ ($f=e$, $i=1$, $j=3$), $\mu\rightarrow e\gamma$ ($f=e$, $i=1,\,j=2$) 
\cite{Masiero:2005ua, Isidori:2013ez,Arana-Catania:2013nha}

\paragraph{LR sector} The terms $(\delta_{LR}^{f})_{ij}^{(Q)}$ which mix left and right sfermions  are given by
\beq
(\delta_{LR}^{f})_{ij}^{(Q)}=\frac{1}{\overline{M}^{2}_{f}}\left(\frac{v_{f}}{\sqrt{2}}(T_{f}^{*})_{ij}-\mu (m_{f}^{*})_i\delta_{ij}\tan\beta^{\pm1}\right),
\eeq
where $v_{f}$, $T_f$, $\mu$ and $m_f$ are Higgses vevs, trilinear terms, $\mu$-term, and quarks/leptons masses respectively. 
Using explicit expressions for soft terms, it is easy to check that $(\delta^{f}_{LR})^{(Q)}$ are suppressed by $m_{f}/\xi$ because trilinear terms $T_{f}$ $\eqref{trilinear}$ are `partially aligned'  to Yukawa matrices $y_{f}$:
\beq\label{dfLR}
|(\delta^{u,d}_{LR})_{ij}^{(Q)}|\approx k_{LR}^{u,d}\frac{|h_{i}^{*}h_{j}|}{2\alpha_{3}^{2}}\frac{(m_{u,d})_{j}}{\xi},\quad |(\delta^{e}_{LR})_{ij}^{(Q)}|\approx k_{LR}^{e}\frac{2|h_{i}h_{j}^{*}|}{\alpha_{2}^{2}}\frac{(m_{e})_{i}}{\xi},
\eeq
where coefficients $k_{LR}^{f}\sim\mathcal{O}(1)$ encompass effect of RG flow from $M_Y$ down to EWSB scale $Q$. Both rough estimate \eqref{dfLR} and numerical scan over $0<h_{1,2}\lesssim0.2$ with $h_3\sim1$ show that 
all 
$(\delta_{LR}^{f})_{ij}$ are at least one order of magnitude below the experimental limits.    

\paragraph{LL squark sector}
It turns out that the experimental data related to the above-mentioned processes also do not significantly constraint $(\delta^{u,d}_{LL})^{(Q)}$.
The reason for that is the following. 
Demanding that the SUSY 1-loop contribution to the meson mass difference $\Delta m_{\mathrm{SUSY}}$ related to $(\delta^{u,d}_{LL})^{(\mathrm{exp})}$ is smaller than the measured value $\Delta m_{\mathrm{exp}}$ results in the following constraint: 
\beq
|(\delta^{u,d}_{LL})_{ij}^{(\mathrm{exp})}|\lesssim c^{u,d}_{ij} f_q(x_q)\frac{M_{\widetilde{g}}}{10^3\,\textrm{GeV}},
\eeq
where $f_q(x_q)=1 + 2(x_q - 1)/5 + 89(x_q - 1)^2/350+27(x_q-1)^3/350+\ldots$ and $x_q=M^2_{\widetilde{g}}/\overline{M}^2_{u,d}$. The coefficients $c^{u,d}_{ij}$ can be expressed in terms of $\Delta m_{\mathrm{exp}}$, meson masses, mesons decay constants and $\alpha_s$ \cite{Gabbiani:1996hi}. Their numerical values are:\footnote{$c^{u}_{13}$ and $c^{u}_{23}$ are not known because mesons containing $t$ quark are not observed in any experiment.}  
$(c^u_{12},c^d_{12},c^d_{13},c^d_{23})=(0.20,0.08,0.20,0.91).$ Hence, one can see that large gluino mass $M_{\widetilde{g}}\approx3.4\cdot10^3\,\textrm{GeV}$ weakens bounds on $(\delta^{u,d}_{LL})^{(\mathrm{exp})}$ such that in the present model one obtains
\beq
\left(|(\delta^{u}_{LL})_{12}^{(\mathrm{exp})}|,|(\delta^{d}_{LL})_{12}^{(\mathrm{exp})}|,|(\delta^{d}_{LL})_{13}^{(\mathrm{exp})}|,|(\delta^{d}_{LL})_{23}^{(\mathrm{exp})}|\right)\lesssim(0.68,0.27,0.68,3.09).
\eeq
Secondly, 
from \eqref{sm-q}:
\beq\label{ddLLij}
(\delta_{LL}^{u,d})_{ij}^{(M_Y)}\approx \frac{h_{i}h_{j}}{640\pi^{2}\alpha_{3}^{2}}(105|h_{3}|^{2}-4\pi(7\alpha_1+45\alpha_2+80\alpha_{3})+\Delta^Q_{ij}).
\eeq
From the previous section, we know that there are two phenomenologically acceptable regions of $h_3$ in the discussed model: $0<h_3<0.25$ and $0.75<h_3<1.0$. Nevertheless, as mentioned in the previous section, we also check the region $0.25<h_3<0.75$ since it can be relevant in some modifications/extensions of the present model. For the further analysis, we choose three values of $h_3$:  $0.1$, $0.5$ and $0.8$, which shall represent those regions. 
  
For $h_3\sim0.8$ there can occur accidental cancellation between terms $\propto |h_3|^2$ and terms $\propto \alpha_r$. In such situation next-to-leading corrections $(\delta^{u,d}_{LL})^{(Q)}$  strongly affect results. For $h_3\sim0.1$ terms $\propto\alpha_r$ dominates.  RG flow gives the following estimate:
\beq
|(\delta^{u,d}_{LL})_{i,j}^{(Q)}|\approx h_ih_j\cdot\left\lbrace
\begin{array}{ccll}
2.5&\textrm{for}&h_3=0.1,\\
1.6&\textrm{for}&h_3=0.5,\\
0.5&\textrm{for}&h_3=0.8,\,h_{1,2}\lesssim0.2.\\
\end{array}\right.
\eeq
Now, it is easy to conclude that eventually 
one gets rather weak constraints on messenger couplings:
\beq
h_1h_2\lesssim
\left\lbrace
\begin{array}{ccll}
0.11&\textrm{for}&h_3=0.1,\\
0.17&\textrm{for}&h_3=0.5,\\
0.54&\textrm{for}&h_3=0.8,\,h_{1,2}\lesssim0.2.\\
\end{array}\right.
\eeq
Numerical scan over $0<h_{1,2}\lesssim0.2$ confirms that meson mixing parameters $\Delta m_{K,B_{d,s}}$ hardly depend  on those messenger couplings. 

\paragraph{RR slepton sector} As one would expect, constraints on $h_{1,2}$ coming from $(\delta^{e}_{RR})^{(Q)}$ are much more restrictive. It turns out that from all the low-energy observables which are sensitive to $(\delta^{e}_{RR})^{(Q)}$ the most constraining is $\mathrm{BR}(
\mu\rightarrow e \gamma)$. 
Similarly to the previous case, after dropping $\propto \{m_f,M_Z\}/\xi$ terms  \eqref{sm-e} yields
\beq
(\delta^{e}_{RR})_{12}^{(M_Y)}\approx\frac{2h_1h_2}{75\pi^2\alpha_2^2}(35|h_3|^2-12\pi(3\alpha_1+5\alpha_2)+\Delta_{12}^E).
\eeq
Numerical tests shows that here cancellation between terms $\propto|h_3|^2$ and terms $\propto\alpha_r$ is not as large as in the previous case i.e. effectively $\Delta_{12}^E$ can be discarded. Again, taking into account RG flow one gets the following effective estimate on the value of $|(\delta^{e}_{RR})_{12}^{(Q)}|$:
\beq\label{deRR12li}
|(\delta^{e}_{RR})_{12}^{(Q)}|\approx h_1h_2\cdot\left\lbrace
\begin{array}{ccll}
25&\textrm{for}&h_3=0.1,\\
3&\textrm{for}&h_3=0.5,\\
15&\textrm{for}&h_3=0.8.\\
\end{array}\right.
\eeq

It is known \cite{Masiero:2005ua,Arana-Catania:2013nha} that when RR sector generates the biggest SUSY 1-loop contribution to $\mu\rightarrow e\gamma$ then
\beq\label{deRR12exp}
|(\delta^{e}_{RR})_{12}^{(\mathrm{exp})}|\lesssim c^e_{12} f_e(x_e)\left(\frac{M_1}{10^3\,\mathrm{GeV}}\right)^2,
\eeq
where $M_1$ is gaugino mass, $f_e(x_e)=1-(x_e-1)/3+22(x_e-1)^2/63-64(x_e-1)^3/189+\ldots$ and $x_e=M_1^2/\overline{M}^2_{e}$. The coefficient $c^{e}_{12}$ can be expressed \cite{Masiero:2005ua,Arana-Catania:2013nha} in terms of muon mass, muon total decay rate, $\alpha_1$, $\sin^2\theta_W$ and experimental limit on $\mathrm{BR}(\mu\rightarrow e\gamma)$. The numerical value of $c^e_{12}$ is $0.038$ and $M_1\approx690\,\mathrm{GeV}$. As a consequence of \eqref{deRR12li} and \eqref{deRR12exp} the upper bound on messenger couplings is
\beq
h_1h_2\lesssim\left\lbrace
\begin{array}{cll}
0.7\cdot10^{-3}&\textrm{for}&h_3=0.1,\\
6.0\cdot10^{-3}&\textrm{for}&h_3=0.5,\\
1.2\cdot10^{-3}&\textrm{for}&h_3=0.8.\\
\end{array}\right.
\eeq
 To verify this result, we used appropriately changed code of \texttt{SPheno-3.3.0}  
and scanned over $0<h_{1,2}\lesssim0.2$ with $h_3$ fixed to either $0.1$ or $0.8$. For each choice of messenger couplings $h_i$ we checked whether the value of $\mathrm{BR}(\mu\rightarrow e\gamma)$ calculated by \texttt{SPheno-3.3.0} is smaller than the current experimental limit $5.7\cdot10^{-13}$ \cite{Adam:2013mnn}. We obtained the following bound on messenger couplings 
for both $h_3=0.1$ and $h_3=0.8$:
\beq\label{hh}
h_1h_2\lesssim0.001.
\eeq
We have to emphasize that  the obtained result   holds for  lepton type Yukawa  couplings. Because we do not expect dramatic differences between leptons  and down-quarks couplings  we expect that a similar bound should hold for the latter case too.

\section{Conclusions}
In this paper, we have presented a model of messenger-flavour mixing through Yukawa couplings based on an assumption of expected hierarchy.
The new couplings has to two major consequences. The largest coupling $h_3$ influences masses of various particles. Detailed analysis shows that there are two windows of acceptable $h_3$
 within the realm of perturbative QFT: 
$0<h_3<0.25$ and $0.75<h_3<1$. The second `window' is what we expect from our model of hierarchical Yukawas.
It appears that masses of the supersymmetric particles are sensitive to values of $h_3$ yielding e.g. very light stop for large $h_3$.
On the other hand mass of the Higgs $h^0$ stays intact.

The model unavoidably leads to non-diagonal soft messes which in consequence trigger FCNC processes. Based on know experimental limits on such processes, we have bounded possible off-diagonal
couplings. It appears that the strongest limits are put by leptons FCNC's.	 We have obtained the following result: 
$h_1 h_2< 10^{-3}$. We can compare this with some existing models of Yukawa couplings. For the assumed lepton-type hierarchy 
$(h_1\sim \ep^4,\ h_2\sim \ep^2,\ h_1\sim 1,\ \ep\sim 0.2)$ one gets that
$h_1 h_2\sim \ep^6\sim 6\cdot 10^{-5}$, while for milder hierarchy of the down-quarks-type $(h_1\sim \ep^2,\ h_2\sim \ep^2,\ h_1\sim 1,\ \ep\sim 0.2)$, $h_1 h_2\sim \ep^5\sim 3\cdot10^{-4}$. Thus, both models of Yukawa couplings are save within experimental bounds.

We hope that the future measurements will let us limits the parameters of the model even farther.

\section*{Acknowledgments}
\vspace*{.5cm}
\noindent The author would like to acknowledge fruitful remarks of Krzysztof Turzy\'nski and Zygmunt Lalak on the manuscript and stimulating discussions with Stefan Pokorski. This work was partially supported by the National Science Center under post-doctoral grant DEC-2012/04/S/ST2/00003.

\end{document}